\documentclass[12pt]{article}

\usepackage{epsf}

\newcommand{\im}{{\rm Im}}
\newcommand{\re}{{\rm Re}}
\newcommand{\beq}{\begin{equation}}
\newcommand{\eeq}{\end{equation}}
\newcommand{\ba}{\begin{array}}
\newcommand{\ea}{\end{array}}
\newcommand{\beqa}{\begin{eqnarray}}
\newcommand{\eeqa}{\end{eqnarray}}
\newcommand{\dis}{\displaystyle}
\newcommand{\cH}{{\cal H}}

\newcommand{\cA}{{\cal A}}
\newcommand{\cO}{{\cal O}}
\newcommand{\br}{{\cal B}}
\newcommand{\cM}{{\cal M}}
\newcommand{\da}{^\dagger}
\newcommand{\no}{\nonumber}
\newcommand{\lsim}{\stackrel{<}{_\sim}}
\newcommand{\gsim}{\stackrel{>}{_\sim}}

\newcommand{\tq}{{\tilde q}}

\newcommand{\td}{{\tilde d}}
\newcommand{\tu}{{\tilde u}}
\newcommand{\tg}{{\tilde g}}

\newcommand{\ttop}{{\tilde t}}

\def\npb#1#2#3{    {\it Nucl. Phys. }{\bf B #1} (#2) #3}
\def\plb#1#2#3{    {\it Phys. Lett. }{\bf B #1} (#2) #3}

\def\prd#1#2#3{    {\it Phys. Rev. }{\bf D #1} (#2) #3}

\def\prl#1#2#3{    {\it Phys. Rev. Lett. }{\bf #1} (#2) #3}
\def\ptp#1#2#3{    {\it Prog. Theor. Phys. }{\bf #1} (#2) #3}

\def\ppnp#1#2#3{   {\it Prog. Part. Nucl. Phys. }{\bf #1} (#2) #3}
\def\rmp#1#2#3{    {\it Rev. Mod. Phys. }{\bf #1} (#2) #3}
\def\zpc#1#2#3{    {\it Z. Phys. }{\bf C #1} (#2) #3}

\def\epjc#1#2#3{   {\it Eur. Phys. J. }{\bf C #1} (#2) #3}
\def\ibid#1#2#3{   {\it ibid. }{\bf #1} (#2) #3}
\def\jhep#1#2#3{   {\it JHEP  }{\bf #1} (#2) #3}

\begin{document}

\thispagestyle{empty}
\begin{flushright}
CERN--TH/2001--355 \\
hep-ph/0112135
\end{flushright}
\vskip 1.0 true cm 

\begin{center}
{\Large\bf $K^+\to \pi^+\nu\bar\nu$: a rising star \\ [5 pt]
 on the stage of flavour physics$^*$} 
 \\ [35 pt]
 {\sc Giancarlo D'Ambrosio${}^{a\dagger}$} and {\sc Gino Isidori${}^{b}$} 
 \\ [20 pt]
{\sl ${}^a$INFN, Sezione di Napoli and Dipartimento di Scienze Fisiche, \\ 
           Universit\`a di Napoli, I-80126 Napoli, Italy} \\ [15 pt]
{\sl ${}^b$Theory Division, CERN, CH-1211 Geneva 23, Switzerland \\ 
 and INFN, Laboratori Nazionali di Frascati, I-00044 Frascati, Italy} 
 \\ [35 pt]

{\bf Abstract}
\end{center}

\noindent
Motivated by the new experimental information reported by the BNL--E787
Collaboration, we analyse the present impact and the future prospects 
opened by the measurement of {$\br(K^+\to\pi^+\nu\bar\nu)$}.
Although still affected by a large error, the  BNL--E787 result 
favours values of $\br(K^+\to \pi^+\nu\bar\nu)$ substantially 
larger than what expected within the Standard Model. As a result, this data 
already provide non-trivial constraints on the unitarity triangle, 
when interpreted within the Standard Model framework.
We stress the importance of the clean relation 
between $\br(K^+\to \pi^+\nu\bar\nu)$,
$\sin2\beta$ and $\Delta M_{B_d}/\Delta M_{B_s}$ 
that in the next few years could provide one of the
deepest probes of the Standard Model in the sector of 
quark-flavour dynamics. 
A speculative discussion about possible non-standard interpretations of a 
large $\br(K^+\to \pi^+\nu\bar\nu)$ is also presented.
Two main scenarios naturally emerge: those with
direct new-physics contributions to the $s \to d \nu\bar\nu$ amplitude 
and those with direct new-physics effects only in $B_d$--$\bar B_d$ mixing. 
Realistic models originating these two scenarios and possible future 
strategies to clearly identify them are briefly discussed.

\vfill 
\noindent 
$^\dagger$ On leave of absence 
at {\it Theory Division, CERN, CH-1211 Geneva 23, Switzerland}.\\
\noindent $^*$ Work supported in part by
TMR, EC--Contract No. ERBFMRX-CT980169
(EURO\-DA$\Phi$NE).

\def\thefootnote{\arabic{footnote}}
\setcounter{footnote}{0}
\setcounter{page}{0}

\newpage
\section{Introduction}
Flavour-changing neutral-current (FCNC) processes 
provide a powerful tool to investigate the flavour 
structure of the Standard Model and its possible
extensions. Among them, $K\to\pi\nu\bar{\nu}$ 
decays are certainly a privileged observatory 
because of their freedom from long-distance uncertainties. 

An important step forward in the difficult challenge
to measure the  $K^+ \to \pi^+ \nu\bar{\nu}$ rate has 
recently been reported by the BNL--E787 Collaboration \cite{E787_new}.
The combined analysis of  BNL--E787 data, including previous
published results \cite{E787_old}, can be summarized as follows:
\beq
\br(K^+\to\pi^+ \nu\bar{\nu}) = \left( 1.57^{~+~1.75}_{~-~0.82} \right) \times 10^{-10}
~ \protect\cite{E787_old}~.
\label{eq:E787}
\eeq
The theoretical estimate of $\br(K^+\to\pi^+ \nu\bar{\nu})$ within the Standard 
Model (SM), as obtained 
by combining the analysis of Ref.~\cite{BB2} with 
an updated Gaussian fit of the Cabibbo-Kobayashi-Maskawa (CKM) matrix \cite{CKM}
(discussed below), reads 
\beq
\br(K^+\to\pi^+ \nu\bar{\nu})_{\rm SM} = \left( 0.72 \pm 0.21 \right) \times 10^{-10}~.
\label{eq:BRSM_val}
\eeq
Although still compatible within the errors, the difference 
between the central values in Eqs.~(\ref{eq:E787}) and (\ref{eq:BRSM_val})
opens interesting perspectives.

The purpose of this letter is twofold. On the one side, we 
analyse the impact of Eq.~(\ref{eq:E787}) within the SM framework: 
as we shall show, despite the large error this result already 
has a non-negligible statistical impact in CKM fits.
We also discuss a possible future strategy to take advantage 
of the theoretically clean nature of $\br(K^+\to\pi^+ \nu\bar{\nu})$, 
$\sin2\beta$ and $\Delta M_{B_d}/\Delta M_{B_s}$. These 
three observables, whose experimental determination  
will substantially improve in the near future, can be combined 
to make one of the most significant tests of the Standard Model
in the sector of quark-flavour dynamics. 

On the other side, we shall discuss possible new-physics scenarios 
that could accommodate a large value of $\br(K^+\to\pi^+ \nu\bar{\nu})$,
assuming that in the future the error in  Eq.~(\ref{eq:E787})
will decrease, without a substantial reduction of the central value.
Interestingly, these scenarios do not necessarily require 
direct new-physics effects in the  $s \to d \nu \bar{\nu}$ 
amplitude: a $\br(K^+\to \pi^+\nu\bar\nu)$ almost twice as 
big as in Eq.~(\ref{eq:BRSM_val}) could also arise with 
direct new-physics effects only in the $B_d$--$\bar B_d$ mixing
amplitude.

\section{$\br(K^+\to \pi^+\nu\bar\nu)$ within the SM}
\label{sect:SM}

Short-distance contributions to the $s \to d \nu \bar{\nu}$ 
amplitude are efficiently described, within the SM, by the 
following effective Hamiltonian \cite{BB2}
\beq
{\cal H}_{eff} = \frac{G_F}{\sqrt 2} \frac{\alpha}{2\pi \sin^2\Theta_W}
 \sum_{l=e,\mu,\tau} \left[ \lambda_c X^l_{NL} + \lambda_t X(x_t) \right]
 (\bar sd)_{V-A}(\bar\nu_l\nu_l)_{V-A}~,
\label{eq:Heff} 
\eeq
where $x_t=m_t^2/M_W^2$, $\lambda_q = V^*_{qs}V_{qd}$ and $V_{ij}$ denote 
CKM matrix elements. The coefficients $X^l_{NL}$ and 
$X(x_t)$, encoding top- and charm-quark loop contributions, 
are known at the NLO accuracy in QCD \cite{BB,MU} and can be found 
explicitly in \cite{BB2}. The theoretical uncertainty in the dominant 
top contribution is very small and it is essentially determined by the 
experimental error on $m_t$. Fixing the $\overline{\rm MS}$ top-quark mass 
to ${\overline m}_t(m_t) = (166 \pm 5)$~GeV we can write
\beq
X(x_t) = 1.51 \left[ \frac{ {\overline m}_t(m_t)}{166~\rm GeV} \right]^{1.15} = 
1.51 \pm 0.05~.
\eeq

The largest theoretical uncertainty 
in estimating $\br(K^+\to\pi^+ \nu\bar{\nu})$ 
originates from the charm sector.  
Following the analysis of Ref.~\cite{BB2},
the perturbative charm contribution is 
conveniently described in terms of the parameter 
\beq
 P_0(X) = \frac{1}{\lambda^4}
\left[\frac{2}{3}X^e_{NL}+\frac{1}{3}X^\tau_{NL}\right] = 0.42 \pm 0.06~.
\label{eq:P0}
\eeq
where $\lambda \equiv |V_{us}|$ is the expansion parameter in 
Wolfenstein's parameterization of the CKM matrix \cite{Wolf}. 
The numerical error in the r.h.s. of Eq.~(\ref{eq:P0})  is obtained from a 
conservative estimate of NNLO corrections \cite{BB2}. 
Recently also non-perturbative effects introduced 
by the integration over charmed degrees of freedom
have been discussed \cite{Falk_Kp}. Despite a precise 
estimate of these contributions is not possible at present
(due to unknown hadronic matrix-elements), these can be 
considered as included in the uncertainty quoted in 
Eq.~(\ref{eq:P0}).\footnote{~The natural order of magnitude 
of these non-perturbative corrections, 
relative to the perturbative charm contribution 
is $m_K^2/(m_c^2 \ln(m^2_c/M^2_W)) \sim 2 \%$.}
Finally, we recall that genuine long-distance effects 
associated to light-quark loops are well below  
the uncertainties from the charm sector \cite{LW}.

With these definitions the branching fraction of $K^+\to\pi^+\nu\bar\nu$ 
can be written as   
\beq
\br(K^+\to\pi^+\nu\bar\nu) = \frac{ \bar \kappa_+ }{ \lambda^2 } ~
\left[ (\im\lambda_t)^2 X^2(x_t) +
\left( \lambda^4 \re\lambda_c  P_0(X)+
       \re\lambda_t X(x_t)\right)^2 \right]~, 
\label{eq:BRSM} 
\eeq
where \cite{BB2}
\beq 
\bar \kappa_+ = r_{K^+} \frac{3\alpha^2 \br(K^+\to\pi^0 e^+\nu)}{
2\pi^2\sin^4\Theta_W} = 7.50 \times 10^{-6} 
\eeq
and $r_{K^+}=0.901$ takes into account the isospin breaking corrections 
necessary to extract the matrix element of the $(\bar s d)_{V}$ current 
from $\br(K^+\to\pi^0 e^+\nu)$ \cite{MP}. 
Employing the improved Wolfenstein decomposition 
of the CKM matrix \cite{Buras_CKM}, Eq.~(\ref{eq:BRSM}) 
describes in the $\bar \rho$--$\bar\eta$ 
an ellipse with small eccentricity,
namely  
\beq
(\sigma\bar\eta)^2 + (\bar \rho- \bar \rho_0)^2 = \frac{ \sigma \br(K^+\to\pi^+\nu\bar\nu) }{ 
\bar \kappa_+  |V_{cb}|^4 X^2(x_t)  }~,
\label{eq:BRSM2} 
\eeq
where 
\beq
\bar \rho_0 =  1 + \frac{\lambda^4 P_0(X)}{|V_{cb}|^2 X(x_t)} \qquad {\rm and} \qquad
\sigma = \left( 1 - \frac{\lambda^2}{2} \right)^{-2}~.
\label{eq:aux1}
\eeq
The ellipse eventually becomes a doughnut once the 
uncertainties on the parameters determining $\bar \rho_0$ 
and on the r.h.s. of (\ref{eq:BRSM2})  are taken into account.

\begin{table}[t]
\caption{Input values used in CKM fits}
\label{tab:inputs}
\begin{center}
\begin{tabular}{|lll|} \hline   
\multicolumn{3}{|c|}{\raisebox{0pt}[15pt][10pt]{Experimental data}} \\ 
\raisebox{0pt}[5pt][10pt]{$\lambda =0.220 \pm 0.002$} & 
$|V_{cb}| = 0.041 \pm 0.002$ & $|V_{ub}/V_{cb}| = 0.085 \pm 0.018$ \\
\raisebox{0pt}[5pt][10pt]{$\Delta M_{B_d} = 0.487  \pm 0.009~{\rm ps}^{-1}$} & 
$\Delta M_{B_s} > 15~{\rm ps}^{-1}$  & $\sin(2\beta) = 0.79 \pm 0.10$ \\ \hline
\multicolumn{3}{|c|}{\raisebox{0pt}[15pt][5pt]{Theoretical inputs}} \\
\raisebox{0pt}[5pt][10pt]{$F_{B_d}\sqrt{{\hat B}_{B_d}} = 230 \pm 40$~GeV}  &
\multicolumn{2}{l|}{$\xi=(F_{B_s}/F_{B_d})\sqrt{{\hat B}_{B_s}/{\hat B}_{B_d}}
\sqrt{M_{B_s}/M_{B_d}} =  1.15 \pm 0.06$} \\
\raisebox{0pt}[5pt][10pt]{${\hat B}_K = 0.8 \pm 0.2$}  & 
$ P_c(\varepsilon) = 0.30 \pm 0.05$  & $ P_0(X) = 0.42 \pm 0.06$
\\ \hline
\end{tabular}
\end{center}
\end{table}

\bigskip

\begin{figure}[t]
\begin{center}
\leavevmode
\epsfysize=9cm
\epsfxsize=15cm\epsfbox{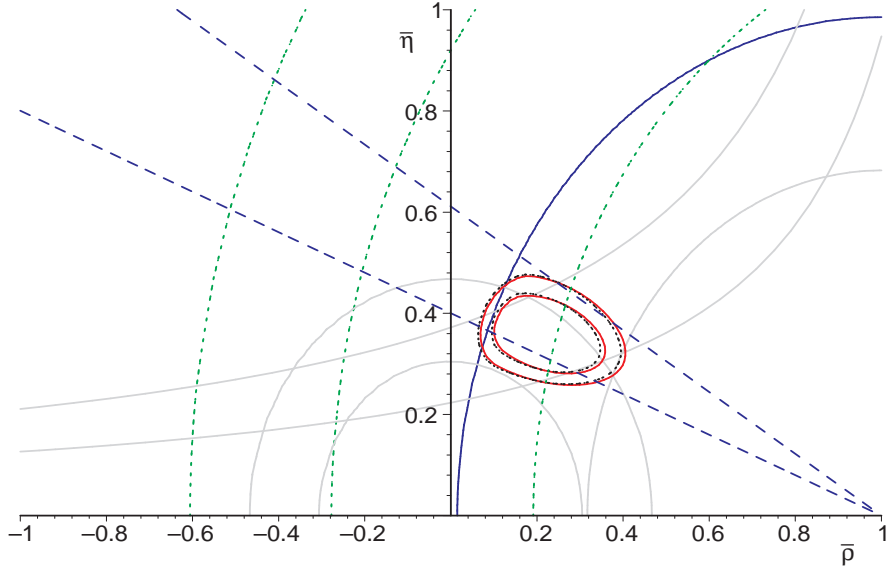}
\vspace{-1.0 cm}
\end{center}
\caption{Global fits in the $\bar\rho-\bar \eta$ plane: the two sets of ellipses
denote $68\%$ and $90\%$ C.L. intervals obtained with (dotted) 
and without (full) the $\br(K^+\to\pi^+ \nu \bar\nu)$ constraint.
The two dotted curves on the left define the 1-$\sigma$ region 
obtained from Eq.~(\ref{eq:BRSM2}), 
setting $\br(K^+\to\pi^+ \nu \bar\nu)$ to the central value 
in (\ref{eq:E787}) and varying all other parameters;
the dotted curve on the right denotes the full 
1-$\sigma$ lower bound imposed by the 
$\br(K^+\to\pi^+ \nu \bar\nu)$ measurement.
The 1-$\sigma$ intervals imposed by $|V_{ub}/V_{cb}|$ (full gray),
$\epsilon_K$ (full gray), $\Delta M_{B_d}$ (full gray), 
$\sin(2\beta)$ (dashed), and 
$\Delta M_{B_d}/\Delta M_{B_s}$ (full dark) are also shown.}
\label{fig:1}
\end{figure}

Stringent bounds about the values of $\bar \rho$ and $\bar \eta$ 
within the SM can be obtained, at present, imposing 
constraints from $|V_{ub}|$, $\Delta M_{B_d}$, $\Delta M_{B_d}/\Delta M_{B_s}$, 
$\epsilon_K$ and $\sin (2\beta)$ \cite{CKMfits}. 
In Fig.~\ref{fig:1} we show the result of a simple Gaussian fit to 
these quantities, using the input values in Tab.~\ref{tab:inputs}: 
all errors have been combined in quadrature, whereas
the $95\%$ upper limit on $\Delta M_{B_s}$ has been treated as an
absolute bound. Up to minor differences [mainly due to the value 
of $B_K$ and the use of $\sin(2\beta)$], the result of 
this fit are in good agreement with more refined analyses available 
in the literature \cite{CKMfits}. The statistical distribution of  
$\bar \rho$ and $\bar \eta$ thus obtained has been used to 
produce the result in Eq.~(\ref{eq:BRSM_val}).
Note that, by construction, the error in Eq.~(\ref{eq:BRSM_val})
does not define a strict interval: it should be interpreted as 
the standard deviation of a Gaussian distribution. 

\begin{figure}[t]
\begin{center}
\leavevmode
\epsfysize=9cm
\epsfxsize=15cm\epsfbox{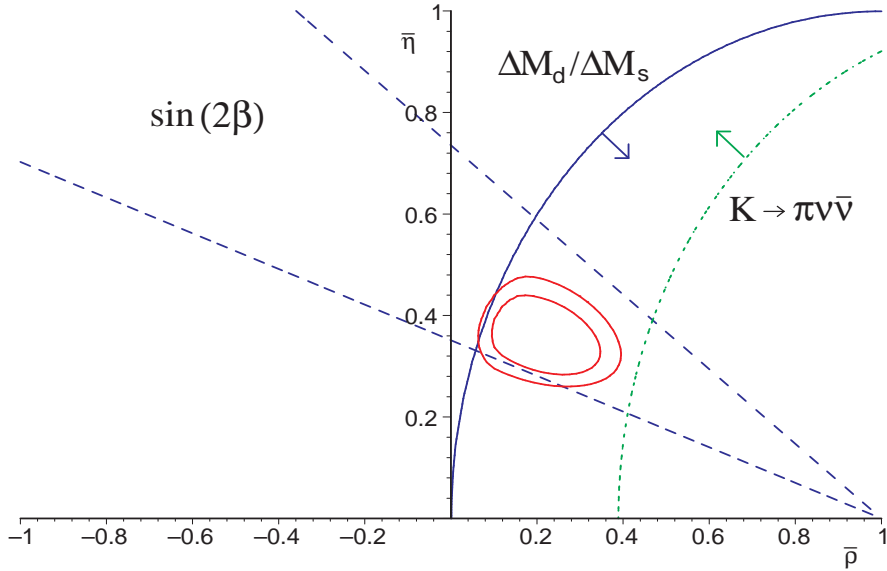}
\vspace{-1.0 cm}
\end{center}
\caption{Allowed region in the $\bar\rho-\bar \eta$ plane using 
only theoretically {\em clean} observables: $90\%$ C.L.
interval imposed by  $\sin(2\beta)$ (dashed); $90\%$ C.L. limit 
from the upper bound on $\Delta M_{B_d}/\Delta M_{B_s}$ (full);
$90\%$ C.L. limit from the lower bound on  
$\br(K^+\to\pi^+ \nu \bar\nu)$ (dotted). For comparison the $68\%$ and $90\%$ C.L.
ellipses from the global fit in Fig.~\protect\ref{fig:1} are also shown.}
\label{fig:2}
\end{figure}

The impact of the present experimental information on  
$\br(K^+\to\pi^+ \nu \bar\nu)$ in the $\bar\rho$--$\bar\eta$ plane
is analysed in Figs.~\ref{fig:1} and \ref{fig:2}. 
Due to the large central value and the non-Gaussian distribution,\footnote{~The 
statistical distribution of $\br(K^+\to\pi^+ \nu \bar\nu)$ has been 
constructed by smooth modification of a Gaussian distribution, fitting the reference 
figures of 68\%, 80\%, 90\% and 98\% C.L. intervals obtained by BNL--E787 
\protect\cite{E787_new}. We are grateful to Steve Kettell for providing us
the reference figures not reported in \protect\cite{E787_new}.}
the BNL--E787 measurement already provides a non-negligible 
statistical input. This is hardly visible in a global fit (Fig.~\ref{fig:1}), 
but is more clear in Fig.~\ref{fig:2}, where the 90\% C.L. 
exclusion limit imposed by the lower bound on 
$\br(K^+\to\pi^+ \nu \bar\nu)$ is shown. 
Due to the large central value,  the overall quality 
of the global fit decreases once the information on $\br(K^+\to\pi^+ \nu \bar\nu)$
is added. However, this is  not a significant 
effect at the moment and,  as shown in Fig.~\ref{fig:2}, 
the SM is still in good shape.

The prediction in Eq.~(\ref{eq:BRSM_val}), based on a global 
CKM fit, suffers to some extent from hadronic uncertainties entering 
the determination of $|V_{ub}|$ and the extraction of $\bar\rho$--$\bar\eta$
constraints from $\epsilon_K$ and $\Delta M_{B_d}$. 
On the other hand, the vertex of the unitarity triangle can in principle 
be determined (up to discrete ambiguities) simply by using 
$\Delta M_{B_d}/\Delta M_{B_s}$ and $\sin(2\beta)$, two quantities 
with a very small theoretical uncertainty. 
By definition, 
\beq
\bar \rho = 1 - R_t \cos\beta~, \qquad \bar \eta = R_t \sin\beta~,
\label{eq:rho_eta}
\eeq
where $R^2_t = (1-\bar\rho)^2 + \bar\eta^2$. 
Expressing $R_t$ as a function of $\Delta M_{B_d}/\Delta M_{B_s}$ \cite{BBL}
we can predict with great accuracy the value of $\br(K^+\to\pi^+ \nu \bar\nu)$ 
in terms of theoretically {\em clean}  observables:
\beqa
\label{eq:master1}
\br(K^+\to\pi^+\nu\bar\nu) &=& \bar\kappa_+  |V_{cb}|^4 X^2(x_t) 
\left[ \sigma R_t^2 \sin^2\beta + \frac{1}{\sigma} 
\left(R_t \cos\beta + \frac{\lambda^4 P_0(X)}{|V_{cb}|^2 X(x_t)}
\right)^2 \right]~, \qquad  \\ 
R_t &=& \frac{ \xi \sqrt{\sigma}}{ \lambda } \sqrt{ \frac{ \Delta M_{B_d}}{ \Delta M_{B_s}} }~
\left[ 1 - \lambda \xi \sqrt{ \frac{ \Delta M_{B_d}}{ \Delta M_{B_s}} }\cos\beta  
+ \cO (\lambda^4) \right]~. \quad
\label{eq:master2}
\eeqa
In the next few years, when the experimental determination of 
$\Delta M_{B_d}/\Delta M_{B_s}$, $\sin(2\beta)$ and 
$\br(K^+\to\pi^+\nu\bar\nu)$ will substantially improve, this 
relation could provide one of the most significant tests of the 
Standard Model in the sector of quark-flavour dynamics. 

Unfortunately at the moment we cannot fully exploit the potential of 
Eqs.~(\ref{eq:master1})--(\ref{eq:master2}) in obtaining a precise prediction of 
$\br(K^+\to\pi^+\nu\bar\nu)$ since $\Delta M_{B_s}$ has not been 
measured yet. Following Ref.~\cite{BB2}, the best we can do at present
is to derive a solid upper bound. Saturating simultaneously the following 
upper limits 
\beqa
\sqrt{ \frac{ \Delta M_{B_d}}{ \Delta M_{B_s}} } < 0.180~, 
\qquad |V_{cb}| < 0.044~, \qquad \sin(2\beta) > 0.5~, \no \\
\xi < 1.3~,  \qquad \qquad P_0(X) < 0.50~, \qquad  X(x_t)< 1.57~, 
\label{eq:assumpt}
\eeqa
that should be regarded as a very conservative assumption, we obtain
\beq
\br(K^+\to\pi^+\nu\bar\nu)_{\rm SM} < 1.32 \times 10^{-10}~.
\label{eq:uppb}
\eeq
By construction it is difficult to assign a probabilistic meaning 
to this result: it should be regarded as an absolute bound
under the assumptions in (\ref{eq:assumpt}). As a consistency check
of this statement, we note that Eq.~(\ref{eq:uppb})
coincides with the $3\sigma$ upper limit derived from the global 
Gaussian fit. We can thus firmly conclude that the central value in 
Eq.~(\ref{eq:E787}) cannot be accommodated within the SM.

\section{New-physics scenarios with a large  $\br(K^+\to\pi^+ \nu \bar\nu)$}
\label{sect:NP}

\begin{figure}[t]
\begin{center}
\leavevmode
\epsfysize=9cm
\epsfxsize=15cm\epsfbox{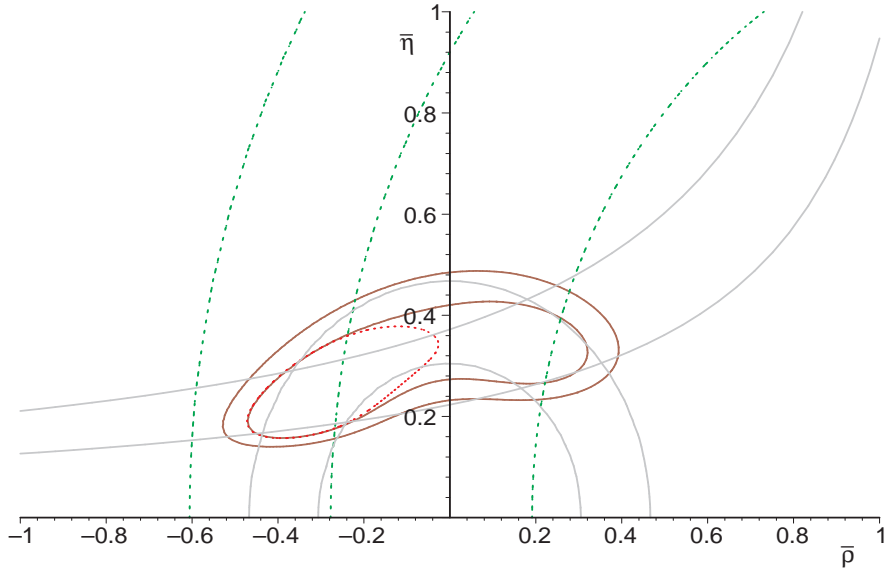}
\vspace{-1.0 cm}
\end{center}
\caption{Allowed region in the  $\bar\rho-\bar \eta$ plane with the 
inclusion of  $\br(K^+\to\pi^+ \nu \bar\nu)$ 
and without $B_d$--$\bar B_d$ data. The two external contours 
denotes $68\%$ and $90\%$ confidence intervals; the inner (dotted) 
one is the $68\%$ confidence interval under the assumption that 
experimental error in (\ref{eq:E787}) is reduced by a factor two.}
\label{fig:NP}
\end{figure}

A stimulating coincidence implied by the experimental result in 
Eq.~(\ref{eq:E787}) is the fact that its central value is well 
in agreement with the constraints imposed by $\epsilon_K$ and $|V_{ub}|$
(see Fig.~\ref{fig:1}). If in the future the error 
on $\br(K^+\to\pi^+ \nu \bar\nu)$ will decrease, without substantial 
changes in the central value, we shall have a conflict only between 
$\br(K^+\to\pi^+ \nu \bar\nu)$ and observables sensitive to 
$B_d$--$\bar B_d$ mixing.
In Fig.~\ref{fig:NP} we show the result of 
a $\bar\rho$--$\bar\eta$ fit without the inclusion of $B_d$--$\bar B_d$ 
data: in this case negative values of $\bar \rho$ are clearly more 
favoured. Remarkably, a similar qualitative indication is obtained 
also by the central values of non-leptonic $B\to K \pi$ decays and, 
in particular, by the deviation of the ratio 
$R_n = \br(B_d \to \pi^\pm K^\mp)/[2\br( B_d \to \pi^0 K^0)]$ 
from one (see Ref.~\cite{BurasFl} and references therein). 
Also in the $B\to K \pi$ case the statistical 
significance of the effect is still quite limited, nonetheless 
there is certainly enough room for speculations about possible 
new-physics effects in $B_d$--$\bar B_d$ mixing. 

As emphasised in the previous section, 
$\Delta M_{B_d}/\Delta M_{B_s}$ and $\sin 2 \beta$ 
on the one side and $\br(K^+\to\pi^+ \nu \bar\nu)$ on the other 
are affected by small theoretical uncertainties, thus 
the potential conflict between $\br(K^+\to\pi^+ \nu \bar\nu)$ 
and $\Delta B=2$ amplitudes is mainly an experimental issue:
if in the future the discrepancy will become more significant it will 
unambiguously signal the presence of new physics. 
Moreover, since the FCNC  $s \to d \nu \bar\nu$ transition  
and $B_d$--$\bar B_d$ mixing both appear only at the loop 
level within the SM, on general grounds both amplitudes
can equally be considered as a good candidates for possible 
non-standard effects. In the following we shall analyse
separately possible new-physics scenarios affecting 
one of these two amplitudes.

\bigskip
\noindent
{\bf Scenario I: non-standard contributions to the   
$s \to d \nu\bar\nu$ amplitude}

\medskip
\noindent
The first question to address about non-standard 
contributions to the observed transition 
$K^+\to \pi^+$ + {\em missing energy} is 
whether the missing energy is due to a $\nu\bar\nu$
pair or not. Since the neutrino pair cannot be
detected, all the information about the decay must
be deduced by the spectrum of the charged pion and 
with only two candidate events this is clearly rather
poor. Nonetheless, some conclusions can already 
be drawn. In particular, we can exclude the possibility that 
these events are generated by a process of the 
type $K^+ \to \pi^+ X^0$, where $X^0$ is a massless particle 
that escapes detection \cite{E787_new,E787_old}.
On the other hand, since $\pi^+$ momenta of the two events are 
almost identical, we cannot exclude yet the possibility 
that these events are due to a two-body decay 
with a massive particle --sufficiently long lived or with 
invisible decay products-- with mass 
$\approx 100$~MeV. This rather exotic scenario could 
easily be discarded in the near future by the observation 
of candidate events with a different kinematical 
configuration. 

\medskip

A general discussions about $K^+\to \pi^+\nu\bar\nu$ beyond the
SM can be found in \cite{GN}. If we assume purely left-handed 
neutrinos and we neglect possible lepton-flavour violations, 
the only dimension-six effective operator relevant to these 
processes is $(\bar sd)_{V}(\bar\nu_l\nu_l)_{V-A}~$  
(as in the SM case) and the measurement of 
$\br(K^+\to \pi^+\nu\bar\nu)$ fix the magnitude 
of its Wilson coefficient. At present 
this is the only available information
about this coefficient, thus there is little we 
can learn from a model-independent analysis. 
The only outcome of such type of analysis 
is an update of the upper bound on $\br(K_L \to \pi^0 \nu\bar\nu)$ \cite{GN},
that in view of Eq.~(\ref{eq:E787}) reads
\beq
\br(K_L\to\pi^0\nu\bar{\nu}) < \frac{ \tau_{K_L} r_{K_L} }{\tau_{K^+}  r_{K^+} }
\br(K^+\to\pi^+\nu\bar{\nu}) < 1.7 \times 10^{-9}~(90\%~{\rm C.L.})~.
\label{GNbd}
\eeq

Among specific new physics models, low-energy supersymmetry 
is certainly one of the most interesting and well-motivated 
scenarios. Supersymmetric contributions to the 
$s \to d \nu\bar \nu$ amplitude have been extensively 
discussed in the recent literature, both within models with 
minimal flavour violation \cite{konig,gambino}
and within models with new sources of 
quark-flavour mixing \cite{NW,BRS,CI}.
As clearly stated in Ref.~\cite{gambino}, minimal models, 
or models without new quark-flavour structures, 
cannot produce a sizeable enhancement of the 
$K^+\to\pi^+\nu\bar{\nu}$ width and would be immediately 
ruled out by a large $\br(K^+\to\pi^+\nu\bar{\nu})$.

Also within models with new sources of quark-flavour 
mixing is not easy to produce sizeable modifications 
of the $s \to d \nu\bar \nu$ amplitude. Excluding 
fine-tuned scenarios with large cancellations 
in $\Delta S=2$ transitions, sizeable
enhancements of $K\to \pi\nu\bar \nu$ rates
can only be generated by chargino-mediated  
diagrams with a large (non-standard) 
${\tilde u}_L^i$--${\tilde u}_R^j$ mixing \cite{NW,BRS,CI}.
Moreover, in the limit of large squark masses 
($M^2_W/M^2_{\tq} \ll 1$) box diagrams 
are systematically suppressed over $Z$-penguin ones 
and can be safely neglected \cite{CI,BCIRS}. 
In this approximation the modification to the SM 
Hamiltonian in (\ref{eq:Heff}) can be obtained by replacing 
$X(x_t)$ with
\beq
X^\prime  = X(x_t) \left[ 1 +  \frac{ A^d_{jl}{\bar A}^s_{ik} F_{jilk} }{8 \lambda_t X(x_t)} 
\right]~,
\label{eq:Zds}
\eeq
where \cite{CI}
\beqa
A^d_{jl} &=& {\hat H}_{l d_L}{\hat V}\da_{1j} 
            - g_t V_{td}{\hat H}_{l t_R}{\hat V}\da_{2j}~, \\ \no
{\bar A}^s_{ik} &=& {\hat H}\da_{s_L k}{\hat V}_{i1} 
            - g_t V_{ts}^*{\hat H}\da_{t_R k}{\hat V}_{i2}~, \\ \no
F_{jilk} &=& {\hat V}_{j1}{\hat V}_{1i}\da~ \delta_{lk}~ k(x_{ik},x_{jk}) 
             - 2 {\hat U}_{i1} {\hat U}\da_{1j}~
\delta_{lk}~ \sqrt{x_{ik}x_{jk}} j(x_{ik},x_{jk}) \nonumber \\
         &&  -\delta_{ij}~ {\hat H}_{k q_L} {\hat H}\da_{q_L l}~ 
k(x_{ik},x_{lk})~. \label{eq:Zds2}
\eeqa
Here $g_t=m_t/(\sqrt{2}M_W\sin\beta)$, ${\hat V}$ and ${\hat U}$
are the unitary matrices that diagonalize the chargino mass matrix
[${\hat U}^* M_\chi {\hat V}\da = \mbox{diag}(M_{\chi_1},M_{\chi_2})$]
and ${\hat H}$ is the one that diagonalizes the up-squark mass matrix
(written in the basis where the $d_L^i-\tu_L^j-\chi_n$ 
coupling is family diagonal and the $d_L^i-\tu_R^j-\chi_n$ 
one is ruled by the CKM matrix). The explicit expressions of 
$k(x,y)$ and $j(x,y)$ can be found in \cite{BRS} and, 
as usual, $x_{ij} = M_i^2/M_j^2$.

The expression (\ref{eq:Zds}) can be further simplified employing 
a perturbative diagonalization of both squark and chargino mass matrices.
In this case we can write 
\beqa
\frac{ X^\prime }{ X(x_t) } &\approx&   
1 + \frac{1}{8 X(x_t)} \left[
 g_t   \frac{ \delta^U_{t_R d_L}\delta^{^\chi}  }{ V_{td} }
 f_1(x_{ij};\tan\beta)
 + g_t \frac{ \delta^U_{s_L t_R} \delta^{^\chi} }{ V_{ts}^* } f_1(x_{ij};\tan\beta) 
  \right. \no\\
&&\qquad\qquad\quad\     \left.
 +  \frac{ \delta^U_{s_L t_R}\delta^U_{t_R d_L} }{ V_{ts}^*V_{td} }  f_2(x_{ij})
 + \cO(V_{ij}^0) \right]~,
\label{eq:Zds3}
\eeqa
where 
\beq
 \delta^U_{t_R q_L} = 
 \frac{({M}^2_{\widetilde U})_{t_R q_L}}{ \langle M_\tq \rangle  }~, 
 \qquad \qquad ~
 \delta^{^\chi} = \frac{ M_W }{ \langle M_\chi \rangle }
 \label{eq:deltas}
\eeq
and $\cO(V_{ij}^0)$ denotes terms not enhanced by 
$V^{-1}_{ts}$ or $V^{-1}_{td}$, which can be safely neglected.
The explicit expressions of the 
adimensional functions $f_{1,2}$, depending on the 
various sparticle mass ratios (and mildly on $\tan\beta$) 
can be extracted from Refs.~\cite{BRS,CI}. 
For $M_{\ttop_R}/M_{\tu_L}\geq 1/2$ and $M_{\chi_j}/M_{\tu_L}\geq 1/3$
one finds $|f_{1}|\lsim 0.1$ and  $|f_{2}|\lsim 0.4$ (the upper figures 
are obtained for the minimal value of $M_{\ttop_R}$). In order to obtain 
$X^\prime/X(x_t) \approx 1.4$, as required by the central value 
in Eq.~(\ref{eq:E787}), the off-diagonal left-right mixing of the squarks 
should satisfy one of the two following conditions: 
\beqa
{\bf\rm i.}  && \left| \delta^U_{s_L t_R}\delta^U_{t_R d_L} \right| \gsim 
\lambda^{-2}  \left| V_{ts}^* V_{td} \right| \no \\ 
{\bf\rm ii.} && \left|  \delta^U_{q_L t_R} \right| \gsim 
\lambda^{-2}  \left| V_{tq} \right| \qquad q=s~{\rm or}~d~.
\label{eq:Zconds}
\eeqa
These requirements are not in contradiction with 
the phenomenological bounds on $\delta^U_{q_L t_R}$ 
imposed by other observables \cite{BCIRS} and are  
consistent with the constraints imposed by the 
stability of the superpotential \cite{Casas}.
However, they necessarily require a rather non-trivial 
structure for the $A$ terms. 

The possible supersymmetric enhancement of the   
$K^+ \to \pi^+  \nu\bar \nu$ rate necessarily 
implies $\cO(1)$ modifications in the 
short-distance $K_L \to \mu^+\mu^-$ amplitude
(sensitive to the real part of the $Z{\bar s}d$ coupling)
and, likely, also $\cO(1)$ effects in 
$\epsilon'/\epsilon$ and $K_L \to \pi^0 \nu \bar\nu$
(sensitive to the imaginary part of the $Z{\bar s}d$ 
coupling). The correlations between these three 
observables have been extensively discussed in Ref.~\cite{BCIRS}.
Unfortunately, the present theoretical uncertainties in 
$K_L \to \mu^+\mu^-$ and $\epsilon'/\epsilon$ and the 
experimental difficulties in the $K_L \to \pi^0 \nu \bar\nu$
case prevent us from drawing definite conclusions 
about the presence of such effects. Concerning 
$K_L \to \mu^+\mu^-$ and $\epsilon'/\epsilon$, 
the situation could possibly improve in the future 
with the help of lattice simulations; however, 
we are clearly quite far from being able to 
detect a $\approx 50\%$ deviation in the pure 
electroweak contribution to these amplitudes.

A rather significant correlation 
can also be established between supersymmetric 
contributions to  $K^+ \to \pi^+  \nu\bar \nu$
and rare FCNC semileptonic $B$ decays.
A general discussion about supersymmetric 
effects in $B \to X_{s,d} \ell^+ \ell^-$ decays,
within the framework of the mass-insertion approximation, 
can be found in \cite{Lunghi}: in the limit where 
only the $\delta^U_{q_L t_R}$ terms are 
substantially different than what expected 
in the minimal scenario, also in these
transitions the main deviations from the SM can be 
encoded in an effective $Z\bar{b}q$ vertex \cite{BHI}.
The smallness of the vector 
coupling of the $Z$ boson to charged leptons
implies that, to a good accuracy,
these effects modify only the Wilson coefficient of the 
axial-current operator
\beq
Q_{10} = \dis\frac{e^2}{4 \pi^2} \bar{q}_L \gamma^\mu b_L 
                   \bar{\ell} \gamma_\mu \gamma_5 \ell~.
\eeq
Employing the same notations as in Eqs.~(\ref{eq:Zds})--(\ref{eq:deltas})
we can write 
\beqa
\frac{C_{10} }{C^{\rm SM}_{10}} &=& 
\left[ 1 +  \frac{ A^q_{jl}{\bar A}^b_{ik} F_{jilk} }{8 \sin^2\Theta_W
V_{tb}^*V_{tq} 
\left| C^{\rm SM}_{10} \right| } 
\right] \no\\
&\approx&      
1 + \frac{1}{8 \sin^2\Theta_W \left| C^{\rm SM}_{10} \right| }
  \left[ g_t   \frac{ \delta^U_{t_R q_L}\delta^{^\chi}  }{ V_{tq} }
  f_1(x_{ij};\tan\beta)
 +  \frac{ \delta^U_{b_L t_R}\delta^U_{t_R q_L} }{ V_{tb}^*V_{tq} }  f_2(x_{ij})
 + \cO(V_{ij}^0) \right]~, \no\\
\label{eq:Zbq}
\eeqa
where $ C^{\rm SM}_{10} \approx -4.2$ \cite{BBL}.
If at least one of the conditions (\ref{eq:Zconds}) is satisfied then 
at least in one of the two cases ($b\to s$ or $b\to d$)
the axial-current operator receives $\cO(1)$ non-standard 
contributions. Since $X(x_t)/(\sin^2\Theta_W | C^{\rm SM}_{10}|) \approx 1.5$,
on general grounds $C_{10}$ is slightly more sensitive to modifications of 
the $Z$-penguin contribution with respect to $X^\prime$.
On the other hand, if $\delta^U_{t_R s_L}$ and 
$\delta^U_{t_R d_L}$ conspire to maximize the effect in Eq.~(\ref{eq:Zds3}),
we can expect a smaller relative impact in Eq.~(\ref{eq:Zbq}).

In the most optimistic case, i.e. in presence of a $100\%$ increase 
of $|C_{10}|$ in the $b\to s$ transition, the effect could possibly
be detected in a short time at $B$-factories, by looking at 
exclusive $B\to (K,K^*) \ell^+\ell^-$ decays. In particular, 
the lepton-forward backward asymmetry in $B\to K^* \ell^+\ell^-$
provides an excellent probe of magnitude and phase of $C_{10}$ \cite{BHI}.
On the other hand, to detect a modification of $C_{10}$ that 
does  not exceed the $30\%$ level in magnitude, either in 
$b\to s$ or in $b\to d$, it is necessary a detailed study of  
inclusive transitions
or pure leptonic decays ($B_{s,d} \to \ell^+ \ell^-$).

Finally, we note that a non-standard $Z\bar{b}q$ vertex 
leads to potentially observable effects also in
inclusive and exclusive $b\to s \nu \bar \nu$ transitions \cite{BHI}. 
In particular, Eq.~(\ref{eq:Zbq}) can trivially be extended 
to the $b\to s \nu \bar \nu$ case with the replacement 
$C_{10} (C^{\rm SM}_{10}) \to  C_{\nu L} (C_{\nu L}^{\rm SM} )$,
where $C_{\nu L}$ is the Wilson coefficient of 
the only dimension-six operator contributing 
to these processes within the SM \cite{BBL}, namely
$(\bar bs)_{V-A}(\bar\nu\nu)_{V-A}~$.

\bigskip
\noindent
{\bf Scenario II: non-standard contributions to
$B_d$--$\bar B_d$ mixing}

\medskip
\noindent
Contrary to the $s\to d \nu\bar\nu$ case, the present 
information about $B$--$\bar B$ mixing 
is already rich and precise.
As a result, the scenario with new physics 
in $B$--$\bar B$ mixing turns 
out to be rather constrained also 
within a model-independent approach. 

The first conclusion that can easily 
be drawn is that this scenario is 
{\em not} flavour blind: we necessarily need to 
modify the SM relation between 
$|V_{td}/V_{ts}|$ and 
$\Delta M_{B_d}/\Delta M_{B_s}$
in Eq.~(\ref{eq:master2}) in order 
to allow a solution with negative $\bar\rho$ 
(see Fig.~\ref{fig:NP}). If new physics 
affects $\Delta M_{B_d}$ and 
$\Delta M_{B_s}$ in the same way,  
with a flavour-blind modification 
of the loop function, then the 
ratio $|V_{td}/V_{ts}|$ extracted from
$\Delta M_{B_d}/\Delta M_{B_s}$ 
would be exactly the same as in the SM.
Since the measurement of $\Delta M_{B_d}$
alone favours positive values of $\bar\rho$ 
(within the SM) and  $\Delta M_{B_s}$ alone is insensitive to  
$\bar\rho$,
the most economical way to implement 
a non-standard scenario with $\bar\rho <0 $ 
is to assume sizeable new-physics effects 
only in $B_d$--$\bar B_d$ mixing.

In presence of new-physics in $B_d$--$\bar B_d$ mixing
we can write, in full generality, 
\beq
\cM^d_{12} ~=~ \frac{1}{2 M_{B_d}} \langle {\bar B}_d^0 | \cH_{\rm eff}^{\Delta B=2} |
 B_d^0 \rangle  ~\propto~ \left[ V_{td}^2  + Z^2 \right]~,
\label{eq:dmz}
\eeq
where $Z^2$ is a complex quantity encoding the 
non-standard contribution, normalized to the SM one
(except for the CKM factor).
The new-physics contribution has been conveniently 
expressed in terms of the square of  
$Z = |Z|e^{i\phi}$ since, in most scenarios, 
contributions to $\Delta B=2$
amplitudes are proportional to the square of some 
$\Delta B=1$ effective coupling. Note, however, that 
$\phi$ is not necessarily the phase of the new 
$\Delta B=1$ effective coupling:
it incorporates also a possible $\pm \pi/2$ 
shift induced by a possible overall minus sign of the 
new contribution with respect to the SM one. 
Denoting by $|V_{td}^0|$ the 
modulus of $V_{td}$ determined by  $\Delta M_{B_d}$ and  
$\Delta M_{B_d}/\Delta M_{B_s}$ assuming only 
SM contributions and by $-\beta_0$ its
phase ($V_{td}^0=|V_{td}^0|e^{-i\beta_0}$) 
determined by $\cA_{\rm CP}(B_d \to \Psi K)$, 
the non-standard contribution $Z$ should satisfy 
the following equation
\beq
|V_{td}|^2 e^{-2i\beta} + |Z|^2 e^{2i\phi} = |V^0_{td}|^2  e^{-2i\beta_0}~,
\eeq
where $V_{td}  = |V_{td}|e^{-i\beta}$ denotes the {\em true} CKM factor.
If we require a solution with $\bar\rho <0$ and $\bar\eta>0$, 
such that $\sin(2\beta) < \sin(2\beta_0)$ and $\cos(2\beta) > 0$,
we then obtain
\beqa
&& 1+\left|\frac{Z}{V_{td}}\right|^4 + 2  \left|\frac{Z}{V_{td}}\right|^2
 \cos(2\phi+2\beta) = \left| \frac{V^0_{td}}{V_{td}} \right|^4 < 1~, 
 \label{eq:condz1} \\
&&  \left|\frac{Z}{V_{td}}\right|^2 \sin(2\phi+2\beta) = 
 \left| \frac{V^0_{td}}{V_{td}} \right|^2 \sin(2\beta-2\beta_0) < 0~.
\label{eq:condz2} 
\eeqa
Even without specifying the exact values of $\bar\rho$ and $\bar\eta$,
the solution of Eqs.~(\ref{eq:condz1})--(\ref{eq:condz2}) requires 
that $(2\phi+2\beta)$ is in the third quadrant, or that $Z$ 
has large imaginary part. Interestingly, this conclusion is 
independent of the discrete ambiguities arising in the 
determination of $\beta_0$ from  $\cA_{\rm CP}(B_d \to \Psi K)$.
If we further impose that 
$\bar\rho$ and $\bar\eta$ are within the inner ellipse in Fig.~\ref{fig:NP},
it is easy to check that $0.7 \lsim |Z/V_{td}| \lsim 1.1 $ 
and $|\phi| \gsim 75^\circ$. From this general analysis 
we conclude that in all models  where the non-standard 
$\Delta B=2$ amplitude is proportional to the square 
of an effective $\Delta B=1$ coupling, the latter 
can be real (i.e. does not imply new CP-violating phases) 
only if there is a relative minus sign between 
SM and non-standard $\Delta B=2$ amplitudes.

\medskip

Similarly to the $s\to d \nu\bar \nu$ case, low-energy 
supersymmetry is one of the most interesting and well-motivated
scenarios to discuss specific predictions. Within the generic 
framework of the mass-insertion \cite{GGMS96} there are several possibilities  
to implement the proper contribution to Eq. (\ref{eq:dmz}). 
For instance, assuming the dominance of gluino box diagrams, 
we can simply adjust the coupling $\delta^D_{b_L d_L}$ 
to produce the desired modification of $B_d$--$\bar B_d$ mixing. 
In particular, $\cO(1)$ corrections are obtained for 
$|\delta^D_{b_L d_L}|\sim \cO(10^{-1})$. Interestingly, 
in this case SM and supersymmetric loop functions 
have the same sign \cite{GGMS96} (assuming $M_\tq \gsim M_\tg$, 
as suggested by RGE constraints \cite{RGEb}) 
thus, according to the general argument discussed above, 
$\delta^D_{b_L d_L}$ must be almost purely imaginary. 

A more specific framework which justifies the existence of 
new flavour structures affecting mainly $B$--physics observables, rather than
$K$ decays or electric dipole moments, is the so-called 
{\em effective supersymmetry} scenario \cite{CKN96}. Within this model
all squarks are rather heavy, with masses of $\cO(10)$ TeV, with the 
exception of left-handed bottom and top squarks, whose masses are kept 
below 1 TeV. By this way supersymmetric contributions to observables 
not involving the third family are naturally suppressed and, 
at the same time, the naturalness problem of the Higgs potential 
is cured by the light squarks of the third family.   
Integrating out the heavy squarks of the first two generations, 
the light sbottom mass eigenstate ($\tilde B$) 
can be written as ${\tilde B} =  Z_{i3} V_{ij} {\td}^j$,
where ${\td}^j$ denote flavour eigenstates, $V$ is the CKM 
matrix and $Z_{ij}$ are coefficients arising by the 
diagonalization of the $3 \times 3$ left-handed 
down-squark mass matrix \cite{CKLN97}. In practice, 
the coupling $(Z_{dB}Z_{bB}^*)$, where 
$Z_{dB}= Z_{i3}  V_{id}$, plays in this context the same 
role as $\delta^D_{b_L d_L}$ in the generic framework of 
the mass insertion approximation. Indeed 
gluino-sbottom box diagrams lead to the following 
effective Hamiltonian \cite{CKLN97}
\beq
{\cal H}_{\rm eff-SUSY}^{\Delta B=2}  =   \frac{\alpha _s ^2}{36 M^2_{\tilde{B}} }  
(Z_{dB}Z_{bB}^*)^2 f(x_{gB})~ \bar{b}_L\gamma_\mu d_L \bar{b}_L \gamma^\mu d_L~,
\label{eq:effH}
\eeq
where\footnote{~Note that there is a missprint in the 
expression of $f(x)$ reported in Ref.~\protect\cite{CKLN97}.}
$f(x) = ( 11+8x-19x^2+26x\ln x+4 x^2 \ln x )/(1-x)^3$.
Employing the reference figure $x_{gB} = 
M^2_\tg/M^2_{\tilde{B}} = 0.1$, it is easy to check that 
\beq
 \left|\frac{Z}{V_{td}}\right| e^{i\phi} \approx  
 \frac{ Z_{dB}Z_{bB}^* }{10^{-2}} ~ \frac{\rm 1~TeV}{ M_{\tilde{B}} }~.
\eeq
The coupling $Z_{dB}Z_{bB}^*$ can naturally be of $\cO(10^{-2})$ \cite{CKLN97},
inducing the desired $\cO(1)$ correction; however, similarly to $\delta^D_{b_L d_L}$, 
also $Z_{dB}Z_{bB}^*$ needs to be almost purely imaginary 
in order to produce the correct sign of the effect 
(i.e. a decrease  of $\Delta M_{B_d}$).   
We further note that a non-trivial flavour structure  
among the first two generations is necessary to ensure that 
$Z_{qB}Z_{bB}^*$ is not proportional to $V_{tq}$ and thus 
the corrections to $\Delta M_{B_s}$ and 
$\Delta M_{B_q}$ are not correlated.

Both within the generic mass-insertion framework and within 
the effective supersymmetry scenario it is not easy to point out
clear correlations between non-standard contributions 
to $B_d$--$\bar B_d$ mixing and those to other observables. 
On the other hand, a clear model-independent indication 
about this non-standard scenario could be obtained 
by a firm experimental evidence (independent from   
$K^+ \to \pi^+ \nu\bar{\nu}$) of $\bar \rho < 0$. 
More precise results on non-leptonic $B\to K \pi$ decays 
would be extremely interesting in this respect \cite{BurasFl}.

\section{Conclusions}

In this letter we have analysed the present impact 
of the new experimental information on 
$\br(K^+\to~\pi^+\nu\bar\nu)$ \cite{E787_new}. Despite an 
apparent large error, the non-Gaussian tail and the large 
central value let us to extract from the BNL--E787 result 
non-trivial constraints on the CKM unitarity triangle.
As we have explicitly shown, the theoretically clean information 
from $\br(K^+\to \pi^+\nu\bar\nu)$, combined with 
$\Delta M_{B_d}/\Delta M_{B_s}$ and $\sin2\beta$, 
already defines a rather narrow region 
in the $\bar \rho$--$\bar \eta$ plane. As emphasised, the precise relation 
linking these three observables will soon provide one of the most 
interesting consistency tests of the Standard Model in the sector of 
quark-flavour dynamics. 

Stimulated by the large central value of the BNL--E787 result, 
we have also presented a speculative discussion about possible 
non-standard interpretations of a large $\br(K^+\to~\pi^+\nu\bar\nu)$.
In general, these can be divided into two big categories: models with 
direct new-physics contributions to the $s \to d \nu\bar\nu$ amplitude 
and models with direct new-physics effects only in $B_d$--$\bar B_d$ mixing. 
In the latter case a large $\br(K^+\to~\pi^+\nu\bar\nu)$ arises 
because of a different CKM fit, which allows a solution with 
$\bar \rho <0$. Supersymmetry with non-minimal flavour structures provides a 
consistent framework to realize both possibilities and, in the 
case of sizeable non-standard contributions to 
$B_d$--$\bar B_d$ mixing, the scenario with heavy masses 
for the first two families emerges as a natural candidate. 

We have outlined the correlations occurring between 
$K^+\to \pi^+\nu\bar\nu$ and rare semileptonic FCNC $B$ 
decays in supersymmetry. If the $s \to d \nu\bar\nu$ amplitude 
receives a sizeable supersymmetric enhancement, a substantial deviation 
from the SM  should be observed either in $b\to d \ell^+\ell^-$ or 
in $b\to s \ell^+\ell^-$ transitions, especially in observables 
sensitive to the axial-current operator $Q_{10}$, such as the lepton 
FB asymmetry.
On the other hand, the smoking-gun for the scenario 
with  new-physics in $B_d$--$\bar B_d$ mixing would be a firm experimental 
evidence of $\bar \rho < 0$, independent from $\br(K^+\to~\pi^+\nu\bar\nu)$,
obtainable for instance by means of $B\to K \pi$ decays.

\section*{Acknowledgements}
We are grateful to Gerhard Buchalla and Gian Giudice 
for many interesting discussions and comments on the
manuscript. We are also in debt with Steve Kettell and 
Laurie Littenberg for informative communications 
about the BNL--E787 experiment.

\end{document}